\documentclass[11pt,twocolumn,twoside,a4paper,amsmath,amssymb,aps,showkeys,showpacs]{revtex4}

\usepackage{amsfonts}
\usepackage{graphics} 
\usepackage{epsfig}
\usepackage{fancyheadings}

\def\cP{{\cal{P}}}
\def\cN{{\cal{N}}}
\def\cO#1{{\cal O}(#1)}
\def\half{{\textstyle \frac12}}

\begin{document}
\thispagestyle{myheadings}
\rhead[]{}
\lhead[]{}
\chead[Dokshitzer]{News of parton dynamics}

\title{A new look for good old parton dynamics}

\author{Yu.L. Dokshitzer}

\affiliation{%
LPTHE, University Paris VI-VII, 4 place Jussieu, Paris, France \\and PNPI, Gatchina, St Petersburg, Russia}%


\begin{abstract}
A short review is given of the idea and of the present status of recently proposed evolution equations that respect the Gribov--Lipatov reciprocity between space-like and time-like parton dynamics in all orders. 
\end{abstract}



\maketitle

\renewcommand{\thefootnote}{\fnsymbol{footnote}}
{\footnotetext[1]{Talk presented at the XXXIX International Symposium on Multiparticle Dynamics, Gomel, Belorussia, September 2009}

\renewcommand{\thefootnote}{\roman{footnote}}


\section{Gribov--Lipatov reciprocity}
\label{sec:GLR}

Evolution equations describing dynamics of distributions of {\em QCD partons}\/  (quarks and gluons) that determine cross sections of hard processes have a long history. 

Logarithmic violation of the Bjorken scaling - dependence of DIS structure functions on the hard scale of the process - and, simultaneously, scale dependence of fragmentation functions in $e^+e^-$ annihilation, were first addressed in a general QFT context by Gribov and Lipatov back in 1971 \cite{GL}. 

These fundamental results were explicitly presented in the language of evolution of a system of partons by Lipatov in 1974 \cite{L}. QCD parton evolution equations followed suit in 1977 \cite{AP,Dok}. 

The universal nature of the parton dynamics goes under the name of
{\em factorisation}\/ of collinear (``mass'') singularities.

Physically, it is due to the fact that quark--gluon multiplication
processes happen at {\em much larger}\/ space--time distances than the
hard interaction itself.  It is this separation that makes it possible
to describe quark--gluon cascades in terms of independent {\em parton
  splitting processes}.  They success one another in a cleverly chosen
{\em evolution time}, $t\sim \ln Q^2$, whose flow ``counts'' basic
parton splittings that occur at well separated, strongly ordered,
space--time scales. 

Perturbative structure of the cross section of a
given process $p$ characterised by the hardness scale $Q^2$ can be
cast, symbolically, as a product (convolution) of three factors (for a
review see~\cite{Guido}):
\begin{eqnarray}\label{eq:01}
  \sigma^{(p)}_h(t) &\!=\!& C^{(p)}[\alpha_s(t)] 
  \otimes \exp\left( \int^t_{t_0} d\tau\, 
    P[\alpha_s(\tau)]\right) \nonumber \\
    & \otimes& w_h(t_0), \qquad t\sim \ln Q^2.
\end{eqnarray}
Here the functions $C[\alpha_s]$ (hard cross section; coefficient
function) and $P[\alpha_s]$ (parton evolution; anomalous dimension
matrix) are perturbative objects analysed in terms of the
$\alpha_s$-expansion.  The last factor $w_h$ embeds non-perturbative
information about parton structure of the participating hadron(s) $h$,
be it a target hadron in the initial state (parton distribution) or a
hadron triggered in the final state (fragmentation function).

Gribov and Lipatov found the splitting functions $P_{AB}(x)$ determining the evolution of parton fluctuations to be {\em identical}\/ for space-like (DIS) and time-like ($e^+e^-$ annihilation) cases:
\begin{eqnarray}\label{eq:ST}
P_{AB}^{(S)}(x_{\mbox{\scriptsize Bjorken}}) &=& P_{AB}^{(T)}(x_{\mbox{\scriptsize Feynman}}), \\ \label{eq:args}
 x_{\mbox{\scriptsize B}} &=& \frac{-q^2}{2(pq)}, \>\> x_{\mbox{\scriptsize F}}\>=\> \frac{2(pq)}{q^2}.
\end{eqnarray}
The identity of ``evolution Hamiltonians'' in the two channels is a highly non-trivial property since the ``energy fraction'' arguments the splitting functions depend on \eqref{eq:args}  are not the same but rather {\em reciprocal}.

Combined together with the Drell--Levy--Yan relation \cite{DLY} (which followed from similarity of Feynman diagrams for the two crossing channels), 
the identity \eqref{eq:ST} translated into an internal symmetry relation for the splitting functions,
\begin{eqnarray}\label{eq:recip}
    P_{AB}(x) = (\mp) x \cdot P_{BA}(x^{-1}), 
\end{eqnarray}
known as the {\em Gribov--Lipatov reciprocity}\/ 
(with the $\mp1$ factor depending on the spins of participating partons $A, B$; for a review see, e.g. \cite{DDT}).

However, the Gribov--Lipatov reciprocity (GLR) was found to be broken beyond the  leading logarithmic approximation (the one-loop parton Hamiltonian). 

Having observed the difference between spase- and time-like two-loop splitting functions, Curci, Furmanski and Petronzio \cite{CFP} have remarked that the violation of the GLR might be of {\em kinematical origin}.

\section{Reciprocity respecting parton evolution equation}

A few years ago the following evolution equation has been suggested in \cite{D,DMS}:
\begin{eqnarray}\label{eq:RREE}
\frac{\partial  D(x,Q^2)}{\partial \ln Q^2}= \int\frac{dz}{z} \cP[z,\alpha_s]\cdot D\left(\frac{x}{z}, z^\sigma Q^2\right)
\end{eqnarray}
where $\sigma=-1(+1)$ for the space-like (time-like) case. Appearance of the combination $z^\sigma Q^2$ in the hardness argument of the parton distribution under integral corresponds to choosing  {\em fluctuation lifetimes}\/ of successive virtual parton states as a logarithmic ordering parameter (``evolution time''). 

Such a choice of an evolution variable is known to be ``wrong'' for either of the two channels. 

Instead, it is the {\em transverse momentum ordering}, $k_\perp^2$ in the S channel, and the {\em angular ordering}, $(k_\perp/z)^2$, for the T-evolution, correspondingly, that make {\em anomalous dimensions}\/ free of series of double logarithmically enhanced terms $(\alpha_s\ln x^2)^n$. Such terms become explosively large in the small-$x$ region which region is practically important both for DIS (high energy scattering) and for jet physics (soft gluon multiplication).     

At the same time, the fluctuation time ordering variable, $k_\perp^2/z$, happens to be lying right in between the two ``clever choices'' and thus preserves the symmetry between the two channels. Therefore, one may hope that formulating parton evolution in terms of equation \eqref{eq:RREE} will result in the universal (channel independent) reciprocity respecting {\em evolution kernel}\/ $\cP(x)$.

{\em Non-locality}\/ of the new equation \eqref{eq:RREE} in
longitudinal ($z$) and transverse variables (hardness scale $Q^2$)
breaks identification of splitting functions with anomalous
dimensions.  

What it offers instead is a {\em link}\/ between the two
channels by means of {\em universal evolution kernel}\/ matrix ${\cal{P}}$, one and the same for T and S evolution.  
In spite of the fact that the new ``splitting functions''
${\cal{P}}$ in \eqref{eq:RREE} do not correspond to any clever choice
of the evolution variable, in either T or S channel (explosive
${\alpha_s\ln^2 x}$ terms being present in both cases), this
universality can be exploited for relating DIS and $e^+e^-$ anomalous
dimensions.

One can expect that by separating the notions of splitting functions
and anomalous dimensions by means of the Reciprocity Respecting
Evolution Equation \eqref{eq:RREE} the Gribov--Lipatov wisdom can
be rescued in all orders.  

This guess was motivated by a remark made by Curci, Furmanski \&\ Petronzio~\cite{CFP} who observed that the GLR violation in the second loop non-singlet quark anomalous dimension amounted to a ``quasi-Abelian'' term $\propto C_F^2$ with a suggestive structure
\begin{eqnarray}\label{eq:CFP}
  \half \left[ P_{qq}^{(2,T)}(x) \!-\! P_{qq}^{(2,S)}(x)\right] 
  &\!\!=\!\!& \int_0^1 \frac{dz}{z} 
  \left\{\! P^{(1)}_{qq}\left(\frac{x}{z}\right) \!\right\}_{\!+}  \nonumber \\
  &\! \cdot \!& P^{(1)}_{qq}(z)\ln z\,.
\end{eqnarray}
Their observation hinted that the GLR violation was not a dynamical
higher order effect but was {\em inherited}\/ from the previous loop
via a non-linear relation \cite{D}.
 
In the Mellin space the convolution \eqref{eq:CFP} translates into
$$P_N\frac{d}{dN}P_N \equiv {P_N\dot{P}_N}.$$  
Let us check that it is this
structure of the GLR breaking that emerges from \eqref{eq:RREE}.

Taking Mellin moments of both sides of the equation we obtain
\begin{equation}\label{eq:shift} 
  \gamma_\sigma(N)\!  D_N(Q^2) \!=\! \int_0^1\!\frac{dz}{z}  z^N P[z, \alpha_s]
  z^{\sigma\partial_{\ln Q^2}} D_N(Q^2) , \nonumber
\end{equation}
where we have used the Taylor expansion trick.  The integral formally
equals
\begin{equation}\label{eq:dshift}
  \gamma_\sigma(N)  \>=\> (D_N)^{-1}\> 
  {\cal{P}}(N+\sigma\partial_{\ln Q^2})\, D_N,
\end{equation} 
expressing the anomalous dimension through the Mellin image of the
evolution kernel with the {\em differential operator}\/ for the argument: 
$$N\to N+\sigma\partial_{\ln Q^2}.$$  
The derivative acts upon
$D_N(Q^2)$ producing, by definition, $\gamma(N) D_N$.  In high orders
it will also act on the {\em running coupling}\/ the anomalous
dimension depends on, $\gamma=\gamma(N,\alpha_s)$. The latter action
gives rise to terms proportional to the $\beta$-function.  Such terms
are {\em scheme dependent}\/ as they can be reshuffled between the anomalous dimension and the coefficient function $C[\alpha_s]$ in the expression \eqref{eq:01}.

Neglecting for the time being such contributions by treating
$\alpha_s$ as constant, \eqref{eq:shift} reduces to a functional equation
\begin{equation}\label{eq:2chan}
  \gamma_{\sigma}(N)  \>=\> {\cal{P}}\left(N+\sigma\gamma_\sigma(N) \right).
\end{equation}
Since $\gamma = \cO{\alpha_s}$, we can expand the argument of the
evolution kernel  perturbatively,
\begin{subequations}\label{gammasol}
\begin{equation}\label{gamser}
{\gamma}_\sigma  \>=\>   
{\cal{P}} \>+\>  \dot{{\cal{P}}} \cdot  \sigma \gamma \>+\>
\half \ddot{{\cal{P}}} \cdot  {\gamma^2} 
\>+\>  \cO{\beta(\alpha)} \>+\> \cO{\alpha^4}.
\end{equation}
Solving \eqref{gamser} iteratively we get
\begin{equation}\label{eq:iter}
  \gamma_\sigma\> =   {\cal{P}} \>+\>  \sigma {\cal{P}} \dot{{\cal{P}}}   \>+\> 
  \left[ {\cal{P}} \dot{{\cal{P}}}^2 + \half {\cal{P}}^2\ddot{{\cal{P}}} \right] +\> \ldots  
\end{equation}
\end{subequations}
Restricting ourselves to the first loop, ${\cal{P}} = \alpha\,
{P}^{(1)}$, with ${P}^{(1)}$ the (Mellin image of) good old LLA
functions, gives
\begin{equation}\label{eq:iter2}
  \gamma_\sigma\> =  \> \alpha {P}^{(1)} \>+\> \alpha^2\, 
  \sigma {P}^{(1)}\dot{P}^{(1)} \>+\>  \ldots  
\end{equation}
The second term on the r.h.s.\ of \eqref{eq:iter2} generates the
two-loop Curci--Furmanski--Petronzio relation \eqref{eq:CFP} all
right. 

Knowing the n-loop anomalous dimension matrix in the S channel, the RREE predicts the anomalous dimensions in the T channel (and vice versa). 

Based on the existing three-loop results for the space-like evolution \cite{MVV}, the corresponding prediction of \eqref{eq:RREE} for the time-like channel was verified in the case of non-singlet anomalous dimensions by Mitov, Moch and Vermaseren in \cite{MiMV}. 

Basso and Korchemsky have traced the origin of the relation \eqref{eq:2chan} to the underlying conformal properties of the theory and described how to embed into the equation effects of non-zero $\beta$-function  \cite{BK}. 

Validity of the Gribov--Lipatov reciprocity for the kernel $\cP$ of the new evolution equation \eqref{eq:RREE} has been checked not only for three-loop non-singlet QCD anomalous dimensions. 

Basso and Korchemsky (in collaboration with Moch) have revisited virtually all known multi-loop QFT results and found GLR to hold for three-loop unpolarized singlet and two-loop polarized QCD distributions (quark transversity, linearly polarized gluon, quark singlet polarized), $\lambda\phi^4$ QFT at four loops, QCD in the $\beta_0\to\infty$ approximation (in all loops),  as well in various SUSY models, including multi-loop anomalous dimensions of the maximally supersymmetric $\cN\!=\!4$ YM theory. In the latter model the GLR was found to hold even in the strong coupling limit, $\alpha\to\infty$  (accessible through the AdS/CFT correspondence).

\bigskip

\section{quasi-elastic limit}

The RREE's first demonstration of force was the derivation of all-order predictions for the structure of subleading singular terms in the expansion of the quark non-singlet and gluon anomalous dimension in the large-$x$ limit \cite{DMS}. 
 
Behaviour of anomalous dimensions in the quasi-elastic ($x\to1$) kinematics can be cast in the following form:
\begin{widetext}
\begin{equation}
\label{largeNx}
\gamma_\sigma(x) =  \frac{A\, x}{(1\!-\!x)_+} + B\, \delta(1\!-\!x)\> +
C_\sigma\ln(1\!-\!x) \> +  D_\sigma + \cO{(1\!-\!x)\log^p(1\!-\!x)}\,,
\end{equation}
\end{widetext}
where the coefficients $A$ and $B$ are the same in the two channels.  

Specific structure of the first --- the most singular --- term
$x/(1-x)$ is dictated by the celebrated Low--Burnett--Kroll (LBK) theorem \cite{LBK}.
It is a consequence of the fact that soft radiation at the level of
$d\sigma\propto d\omega(\omega^{-1}+\mbox{const})$ has {\em classical
  nature}.  
  
  The coefficient $A$ in front of this structure has a
meaning of the ``physical coupling'' measured by the intensity of
relatively soft gluon emission. This coefficient (calculated in three
loops in the MS-bar scheme) is known to universally appear in all
observables sensitive to soft gluon radiation: quark and gluon Sudakov
form factors and Regge trajectories, threshold resummations, singular
part of the Drell--Yan $K$-factor, distributions of jet event shapes
in the near-to-two-jet kinematics, heavy quark fragmentation
functions, etc. The structure \eqref{largeNx} applies to the
large-$x$ behaviour of the $g\to g$ anomalous dimension as well, with
$A_{(g)}/A_{(q)}=C_A/C_F$, {\em in all orders}.

Quantum effects show up only at the level of $d\sigma\propto \omega\,
d\omega$ that is at the level of contributions that were {\em
  neglected}\/ in \eqref{largeNx}.  At one loop, subleading terms
$C\ln(1-x)$ and the constant $D$ in \eqref{largeNx} are absent.
This suggests that in higher loops they should emerge as ``inherited''
rather than non-trivial entries.

Indeed, to keep under control all the terms in \eqref{largeNx} it
suffices to use the $x\to1$ 
asymptote of the one-loop evolution kernel ${\cal{P}}^{(1)}=P^{(1)}$ to derive from RREE \eqref{eq:RREE}
{\em all-order relations}~\cite{DMS}
\begin{subequations}
\begin{eqnarray}
\label{eq:C}
 C_\sigma\>&=&\> -\sigma\, A^2, \\
\label{eq:D}
 D_\sigma\>&=&\> -\sigma\,A\,B.
\end{eqnarray}
\end{subequations}
The relation \eqref{eq:C} is ``conformal'' while \eqref{eq:D} acquires
correction due to running of the coupling~\cite{BK}. 

The ideas of the universal evolution equation have recently found an interesting application in the work by Laenen, Magnea and Stavenga who have employed the RREE as means of improving threshold resummations \cite{LMS}.

\section{$\cN\!=\!4$ SYM}
\label{sec:N4SYM}

QCD shares the gluon sector with supersymmetric YangÐMills models (SYM). 
This suggests to explore supersymmetric partners of QCD in order
to shed light on the subtle structure of the perturbative quarkÐgluon dynamics.

QCD is not an integrable quantum field theory. In spite of this, in certain sectors
of the chromodynamics the integrability does emerge \cite{integr}. This happens, markedly, in the problem of high energy Regge behaviour of scattering amplitudes in the large-$N_c$ approximation (planar 't Hooft limit), in the spin 3/2 baryon wave function, for the scale dependence of specific (maximal helicity) quasi-partonic operators (for review see \cite{integr2}). 
What all these problems have in common, is the irrelevance of quark degrees of freedom and the dominance of the classical part of gluon dynamics, in the sense of the LBK 
theorem \cite{LBK}.

The higher the symmetry, the deeper integrability. The maximally supersymmetric $\cN\!=\!4$ YM theory is exceptional in this respect. A string of recent theoretical developments \cite{integr2,BS,Min} hinted at an intriguing possibility that this QFT, super-conformally invariant at the quantum level ($\beta(\alpha)\equiv 0$), may admit an {\em all loop solution}\/ for anomalous dimensions of its composite operators. 

The $\cN\!=\!4$ SYM being an integrable model, there exists a powerful technology based on the Bethe Ansatz Equations well suited for perturbative calculation of anomalous dimensions of composite operators to multi-loop accuracy. The so called ``universal anomalous dimension'' of the {$\cN\!=\!4$} SYM theory is given by the ``maximal transcedentality'' Euler--Zagier harmonic sums \cite{KL,KLOV}. 

Applied to this theory, the kernel of \eqref{eq:RREE} was found to respect GLR in {\em four loops}\/ for the leading twist two \cite{Bec4}, as well as in {\em four}\/ \cite{tw3} and {\em five loops}\/ \cite{Bec5} for twist three operators. 

The  RREE was found to significantly simplify the structure of high order terms in the ``universal anomalous dimension'' of the theory \cite{DM}. 

Given that in the leading order evolution kernel of the $\cN\!=\!4$ SYM is {\em purely classical}\/ in the LBK sense,
\begin{equation}  
\cP^{(1)}(x)\>=\> \frac{x}{1-x} \>+\> \mbox{no quantum corrections}, 
\end{equation}
one may hope to derive one day a {\em one-line-all-loops}\/ expression for the anomalous dimension of this theory, in which higher order terms are  dynamically ``inherited'' from the first loop. 

QCD would greatly benefit from such a solution, since this $\cN\!=\!4$ SYM finding would put under full theoretical control the {\em dominant part}\/ of the perturbative QCD gluon dynamics.

\label{last}
\end{document}